\documentclass[journal=jacsat,manuscript=article]{achemso}
\setkeys{acs}{articletitle=true}

\usepackage[version=3]{mhchem} 

\newcommand{\D}{{\rm d}}
\newcommand{\br}{{\bf r}}
\newcommand{\mint}[1]{\int\! \D #1 \, }

\author{Diogo A. F. Almeida}
\email{diogo.almeida@student.fisica.uc.pt}
\affiliation{CFisUC, Department of Physics, University of Coimbra, Rua Larga, \mbox{3004-516} Coimbra, Portugal.}
\author{Micael J. T. Oliveira}
\email{micael.oliveira@mpsd.mpg.de}
\affiliation{Max Planck Institute for the Structure and Dynamics of Matter, Luruper Chaussee 149, D-22761 Hamburg, Germany.}
\author{Bruce F. Milne}
\email{bfmilne@uc.pt}
\affiliation{CFisUC, Department of Physics, University of Coimbra, Rua Larga, \mbox{3004-516} Coimbra, Portugal.}

\title[\texttt{achemso} Bismuth carbide clusters]
{First-principles characterisation of spectroscopic and bonding properties of cationic bismuth 
carbide clusters}

\begin{document}
\begin{abstract}
Vibrational and electronic absorption spectra calculated at the (time-dependent) density functional theory level for the bismuth carbide clusters \ce{Bi_{n}C_{2n}^+} ($3 \le n \le 9$) indicate significant differences in types of bonding that depend on cluster geometry. Analysis of the electronic charge densities of these clusters highlighted bonding trends consistent with the spectroscopic information. The combined data suggest that larger clusters ($n > 5$) are likely to be kinetically unstable in agreement with the cluster mass distribution obtained in gas-aggregation source experiments. The spectral fingerprints of the different clusters obtained from our calculations also suggest that identification of specific \ce{Bi_{n}C_{2n}^+} isomers of should be possible based on infra-red and optical absorption spectroscopy.
\end{abstract}

\section{Introduction}

Metal carbides (MC's) have come under increasing scrutiny in recent years due to the novel properties possessed by many members of this chemical family. The great majority of the MC's studied to date have been cluster-type structures involving transition metals.\cite{Oyama1992,Oyama1996,Bernstein2013,Yuan2014} Transition metal carbides (TMC's) especially have received considerable attention and are being investigated for applications in areas such as catalysis in chemicals production and fuel cell design.\cite{Delannoy2000,Ganesan2005,Liu2013}

Whilst TMC's have been the subject of study for a number of decades, in recent years interest has begun to turn towards the later \emph{p}-block metallic elements.\cite{Bernstein2015,Tamura2016} These elements offer not only chemistries that differ greatly from the transition metals, but can also be expected to display unique properties due to strong relativistic effects resulting from the presence of very heavy atomic nuclei.

Relatively early examples of this class of main group metal carbides are the range of cationic cage-like bismuth carbide metalcarbohedrons \ce{Bi_{n}C_{2n}^+} that were synthesised using a gas-aggregation source by Yamada and Nakagawa (2009).\cite{Yamada2009} The creation of metal carbides containing such a heavy element is of considerable interest as it provides an opportunity to study the properties of (\emph{p}-block) metal carbides taken to the extremes of difference in masses between their metal and carbon constituents.

The characterisation of the bismuth carbides thus produced involved mass spectroscopic analysis of the resulting gas stream combined with density functional theory (DFT) calculations of the relative energies of possible structural isomers. Whilst this permits a tentative prediction of the species actually produced by the gas-aggregation method, a more complete spectroscopic analysis would be required in order to positively identify the cluster/cage geometries.

We present here theoretical calculations of vibrational and optical absorption spectra for the various structural isomers posited by Yamada and Nakagawa. Spectroscopic features that will permit clear characterisation of these cluster types have been identified and suggest fundamental differences in the bonding displayed by these clusters and, in particular, within \ce{C_{2}} dicarbon units. Analysis of the properties of the charge density of these clusters support the spectroscopic observations regarding bonding and help to rationalise the observed relative yield of clusters of differing masses.

\section{Methods}

\begin{figure}[!h]
\centering
\includegraphics[scale=0.7]{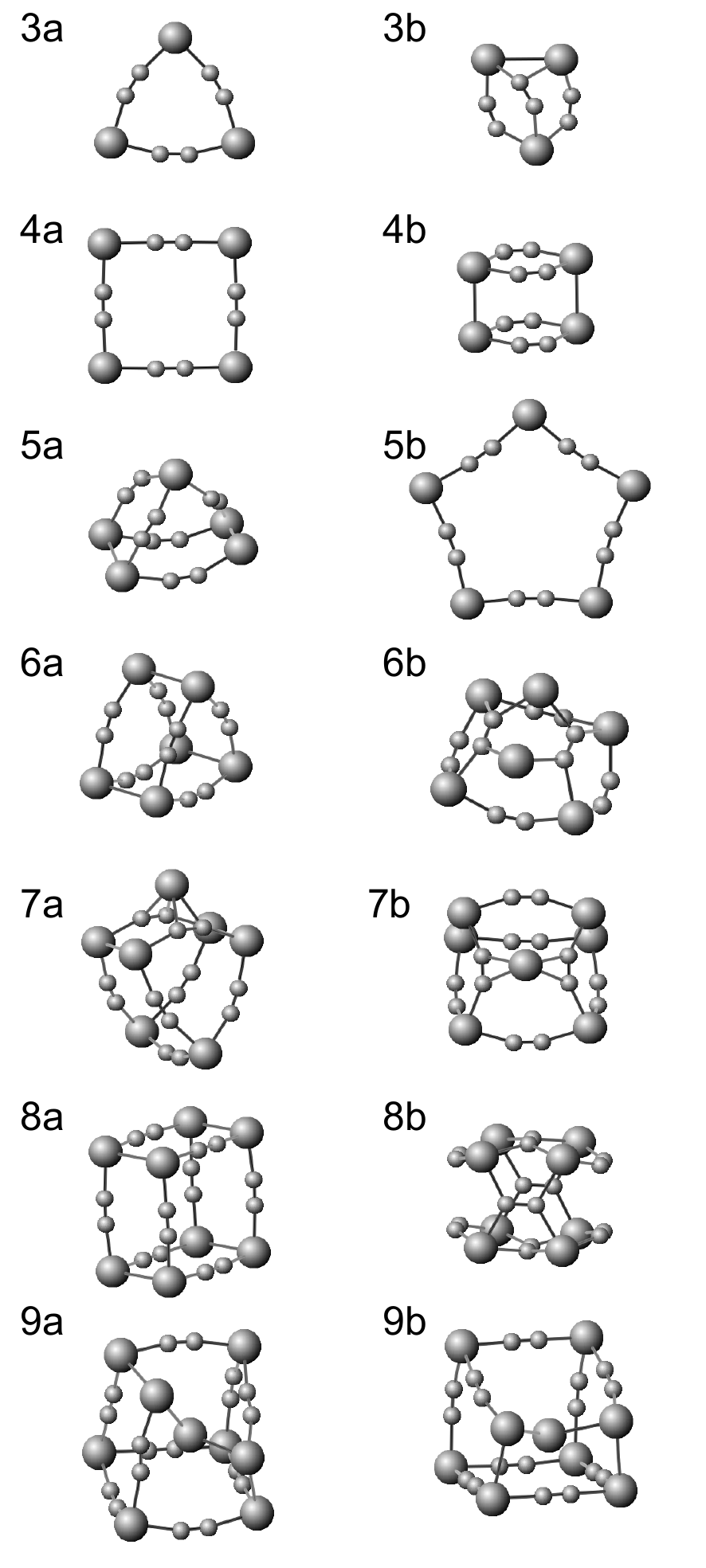}
\caption{Bismuth carbide cations \ce{Bi_{n}C_{2n}^+} ($3 \le n \le 9$) 
         investigated in this work. Geometries optimised at the DFT/B3LYP/Def2-TZVP 
         level. Large spheres = bismuth, small spheres = carbon. Molecular graphics 
         created using Marvin (www.chemaxon.com).}
\label{fig:structures}
\end{figure}

The systems chosen for study in the present work were the \ce{Bi_{n}C_{2n}^+} ($3 \le n \le 9$) cations suggested previously as candidates for the species detected in the gas aggregation source experiment by Yamada and Nakagawa (Fig. \ref{fig:structures}).\cite{Yamada2009} The smaller clusters with $n = 1$ or $n = 2$ were not observed experimentally and were therefore not included in the present study. Similarly, no clusters with $n > 9$ were detected in the gas aggregation experiments and for this reason the maximum number of Bi atoms per cluster was limited to 9 in the current work. The numbering system used in the present work was kept the same as that in the paper by Yamada and Nakegawa in order to facilitate comparison with the original article.

\begin{table}[!h]
\small
 \caption{\ Spin multiplicities of the \ce{Bi_{n}C_{2n}^+} clusters studied in the present work}
  \begin{tabular*}{0.48\textwidth}{@{\extracolsep{\fill}}ccc}
    \hline
     \emph{n} & a & b \\
    \hline
    3 & triplet & singlet \\
    4 & quartet & doublet \\
    5 & singlet & singlet \\
    6 & doublet & doublet \\
    7 & singlet & singlet \\
    8 & doublet & doublet \\
    9 & singlet & singlet \\
    \hline
  \end{tabular*}
  \label{tbl:multiplicities}
\end{table}

Geometry optimisations and vibrational analyses/calculations of infra-red (IR) spectra were performed at the DFT/B3LYP level\cite{Becke1993,Stephens1994} using GAMESS-US (11 August 2011 (R1) release).\cite{Schmidt1993} The polarised \mbox{triple-$\zeta$} \mbox{Def2-TZVP} basis set\cite{Weigend2005} was used in all B3LYP calculations. Scalar relativistic effects were included in the B3LYP/Def2-TZVP calculations through the use of effective core potentials (ECP) for Bi atoms.\cite{Metz2000} Spin multiplicities used (the same as those in the article by Yamada and Nakagawa) are shown in Table \ref{tbl:multiplicities}.\cite{Yamada2009} Relative energies for the 'a' and 'b' forms of each \ce{Bi_{n}C_{2n}^+} cluster were calculated including a correction for zero-point energy (ZPE). The B3LYP/Def2-TZVP ZPE was scaled by 0.965 in line with the data contained in the NIST CCCBDB database (http://cccbdb.nist.gov/). The IR spectra shown in Figure \ref{fig:IR} were processed using the Gabedit software package in order to visualise the calculated vibrational modes and to fit Lorentzian curves to the energies and intensities calculated with GAMESS-US.\cite{Allouche2011}

The optical absorption cross-sections were calculated using time dependent 
density functional theory (TDDFT).~\cite{rg84:997,mmngr2012} In particular, we 
used the real-time method as implemented in the {\sc octopus} code.~\cite{caoralmgr06:2465,aasoncmalarm12:233202,C5CP00351B} In this method, the charge density $\rho(\br)$ of the 
system 
is perturbed by a dipolar electric field
\begin{equation}
 \delta v(\br, t) = - \mathbf{E}\cdot\mathbf{r} \delta(t)\
\end{equation}
that excites all the frequencies of the system equally. The response of the 
system, as characterized by the induced dipole moment
\begin{equation}
 \delta \mathbf{p}(t) = \mint{\mathbf{r}} \left[ \rho(\br, t) - \rho(\br, 
0)\right] \, \br \, ,
\end{equation}
is then obtained by solving numerically the time-dependent Kohn-Sham equations. 
This allows for the determination of the dynamical polarisability of the system
\begin{equation}
 \alpha_{ij}(\omega) = \frac{\delta p_i(\omega)}{E_j},
\end{equation}
which in turn is trivially related to the optical absorption 
cross-section $\sigma(\omega)$
\begin{equation}
\sigma(\omega) = \frac{4\pi\omega}{c} \mathrm{Im}\left\lbrace \left\langle 
\alpha(\omega) \right\rangle \right\rbrace\,,
\end{equation}
where the brackets denote orientational averaging.

In {\sc octopus} the core-electrons are replaced by norm-conserving pseudopotentials. Since relativistic effects come mainly from the core electrons, these have to be taken into account through the pseudopotential. Therefore, a fully relativistic pseudopotential was generated for bismuth using the relativistic extension of the Troullier-Martins scheme~\cite{ehv01:125121} as implemented in the Atomic Pseudopotentials Engine package.~\cite{on08:524} This pseudopotential is in turn used in {\sc octopus} in the form of $j$-dependent Kleinman-Bylander projectors,~\cite{th01:073106} such that it is possible to turn the spin-orbit coupling term on and off at will.~\cite{onmr09:214302} Because relativistic effects are much smaller in carbon, they can safely be neglected when generating the corresponding pseudopotential and in this case the usual non-relativistic Troullier-Martins scheme~\cite{tm91:1993} was used instead.

{\sc octopus} is a real-space finite-differences code where most of the relevant quantities are represented in a regular rectangular grid. Therefore, the numerical convergence of the calculated quantities depends on two things: the spacing of the grid and the size of the simulation box. We found that a spacing of 0.22 \AA\ and a simulation box constructed from a union of spheres of radius 6.1 \AA\ centred at the nuclei was needed to converge the position of the optical absorption peaks within 0.1 eV.

All TDDFT calculations were performed using the Perdew-Burke-Ernzerhof (PBE) generalized gradient correction (GGA) for the exchange-correlation functional and used the non-collinear spin approach for the incorporation of spin-orbit coupling effects.~\cite{pbe96:3865} The magnetic moments obtained from these relativistic PBE calculations are shown in Table \ref{tbl:magnetic_moments} and are systematically lower than what would be expected from the spin multiplicities employed in the scalar relativistic B3LYP geometry optimisations. This is caused by the spin-orbit coupling, which induces non-collinearity in the magnetization density. We recover the same total magnetic moments as in the scalar relativistic B3LYP geometry optimisations by switching off the spin-orbit coupling in the PBE calculations.

\begin{table}[!t]
\small
\caption{\ Magnetic moments of the \ce{Bi_{n}C_{2n}^+} clusters obtained from relativistic PBE calculations incorporating spin-orbit effects. Magnetic moment obtained without incorporation of spin-orbit effects are given in parentheses.}
  \begin{tabular*}{0.48\textwidth}{@{\extracolsep{\fill}}ccc}
    \hline
     \emph{n} & a & b \\
    \hline
    3 & 1.77 \, (2.00) & 0.00 \, (0.00) \\
    4 & 1.12 \, (3.00) & 0.88 \, (1.00) \\
    5 & 0.00 \, (0.00) & 0.00 \, (0.00) \\
    6 & 0.90 \, (1.00) & 0.23 \, (1.00) \\
    7 & 0.00 \, (0.00) & 0.00 \, (0.00)\\
    8 & 0.90 \, (1.00) & 0.96 \, (1.00) \\
    9 & 0.00 \, (0.00) & 0.00 \, (0.00) \\
    \hline
  \end{tabular*}
  \label{tbl:magnetic_moments}
\end{table}

Investigation of the bonding interactions in the bismuth carbide clusters was performed based on the analysis of the charge density $\rho(\br)$ and used two approaches. Covalent interactions were investigated using the electron localisation function (ELF)\cite{Becke1990,Silvi1994,Kohout2002}

\begin{equation}
\mathrm{ELF} = \frac{1}{1 + \left(\frac{D_{\sigma}}{D^0_{\sigma}}\right)^2}\,,
\end{equation}

\noindent which provides a measure of the deviation of the actual charge density of the system under study from that of a homogeneous electron gas of the same density, thus acting as an indicator of the degree of localisation of electron density at a given point. The ELF has the property that $0 \leq \mathrm{ELF} \leq 1$ with ELF = 1 corresponding to full localisation and ELF= 0.5 being equal to the electron gas. ELF calculations were performed using the Gabedit software version 2.5.0.\cite{Allouche2011}

Non-covalent interactions (NCI) were analysed using a new implementation of the NCI index of  Johnson et al.\cite{Johnson2010} PyNCI is maintained by one of the authors (DAFA) and is available to download from https://gitlab.com/diofalmeida/pynci.\cite{Almeida2020} 

Central to the NCI index method is the reduced density gradient (RDG), $s(\br)$

\begin{equation}
s(\br) = \frac{1}{2{(3\pi^2)}^{1/3}} \frac{| \nabla \rho(\br) |}{\rho(\br)^{4/3}}\,,
\end{equation}

\noindent which is obtained from the charge density and has large positive values in low density regions such as those far from the molecule, but importantly has values approaching zero in regions of space contributing to (non)bonding interactions between atoms. 

An additional component in the NCI method is the second eigenvalue ($\lambda_2$) of the Hessian matrix of the electron density, $\nabla^2 \rho(\br)$. This has the property that, in inter-atomic regions where $s(\br)$ is small, $\lambda_2 < 0$ corresponds to a stabilising or bonding interaction, whereas if $\lambda_2 > 0$ the interaction is destabilising or non-bonding. When combined as 2D scatter plots or as 3D spatial plots of the RDG isosurface coloured by the value of $\lambda_2$ these two properties provide a powerful insight into the bonding characteristics of the regions of charge density involved in non-covalent interactions. 

The electronic charge densities used in the ELF/NCI analysis were calculated at the all-electron level with the SCAN meta-GGA functional\cite{Sun2015, Perdew2016} using the B3LYP/Def2-TZVP geometries. Scalar relativistic effects were incorporated with the second-order Douglas-Kroll-Hess (DKH2) method\cite{Douglas1974, Hess1985, Hess1986, Jansen1989, Wolf2002} and the spin-orbit contribution was calculated using the one-centre spin-orbit mean field (SOMF(1X))\cite{Neese2005} approach. Basis sets re-contracted for use in relativistic calculations were employed (DKH-Def2-TZVP\cite{Buhl2008} for carbon and SARC-DKH-TZVP\cite{Pantazis2012} for bismuth). Coulomb integrals were approximated with the resolution of the identity approach using the Def2/J and SARC/J auxiliary basis sets.\cite{Neese2003, Weigend2006} Electronic charge densities were output in Gaussian cube format for analysis with the PyNCI program. All SCAN calculations were performed with the Orca software package version 4.2.1.\cite{Neese2012}

\section{Results and discussion}

\subsection{Relative energies of bismuth carbide isomers}

\begin{table}[!t]
\small
  \caption{\ Relative B3LYP energies of \ce{Bi_{n}C_{2n}^+} clusters including ZPE correction (eV). Previous LDA/TZP values of Yamada and Nakagawa included for comparison.\cite{Yamada2009}
  }
  \begin{tabular*}{0.48\textwidth}{@{\extracolsep{\fill}}ccccc}
    \hline
     & \multicolumn{2}{c}{B3LYP/Def2-TZVP} & \multicolumn{2}{c}{LDA/TZP\cite{Yamada2009}}\\
     \emph{n} & a & b & a & b\\
    \hline
    3 & 0 & 0.97 & 0 & 0.84 \\
    4 & 0 & 1.78 & 0 & 1.55 \\
    5 & 0 & 0.72 & 0 & 0.48 \\
    6 & 0 & 1.14 & 0 & 0.71 \\
    7 & 0 & 0.13 & 0 & 0.05 \\
    8 & 0 & 3.12 & 0 & 1.72 \\
    9 & 0 & 0.28 & 0 & 0.34 \\
    \hline
  \end{tabular*}
  \label{tbl:energies}
\end{table}

The relative energies of the \ce{Bi_{n}C_{2n}^+} clusters obtained in the present work (see Table \ref{tbl:energies}) agree well with those obtained in the original article by Yamada and Nakagawa. In all cases the 'a' cluster is found to be the most thermodynamically stable of the two alternatives. With the exception of the \ce{Bi_{9}C_{18}^+} cluster, the energy differences were all larger than those calculated at the LDA/TZP level, however, the general trend remained the same. It seems reasonable to assume that the values obtained in the present work are likely to be more accurate since the B3LYP functional is known to yield superior thermochemical energies to those obtained with LDA. Similarly, although the basis sets used in this and the previous work are of similar valence-space quality, both being triple-$\zeta$, the Def2-TZVP set contains several polarisation functions per atom whereas the TZP basis set contains only one polarisation function per atom making it less flexible.

\subsection{Vibrational spectra}

\begin{figure}[]
\centering
\includegraphics[width=8.5cm]{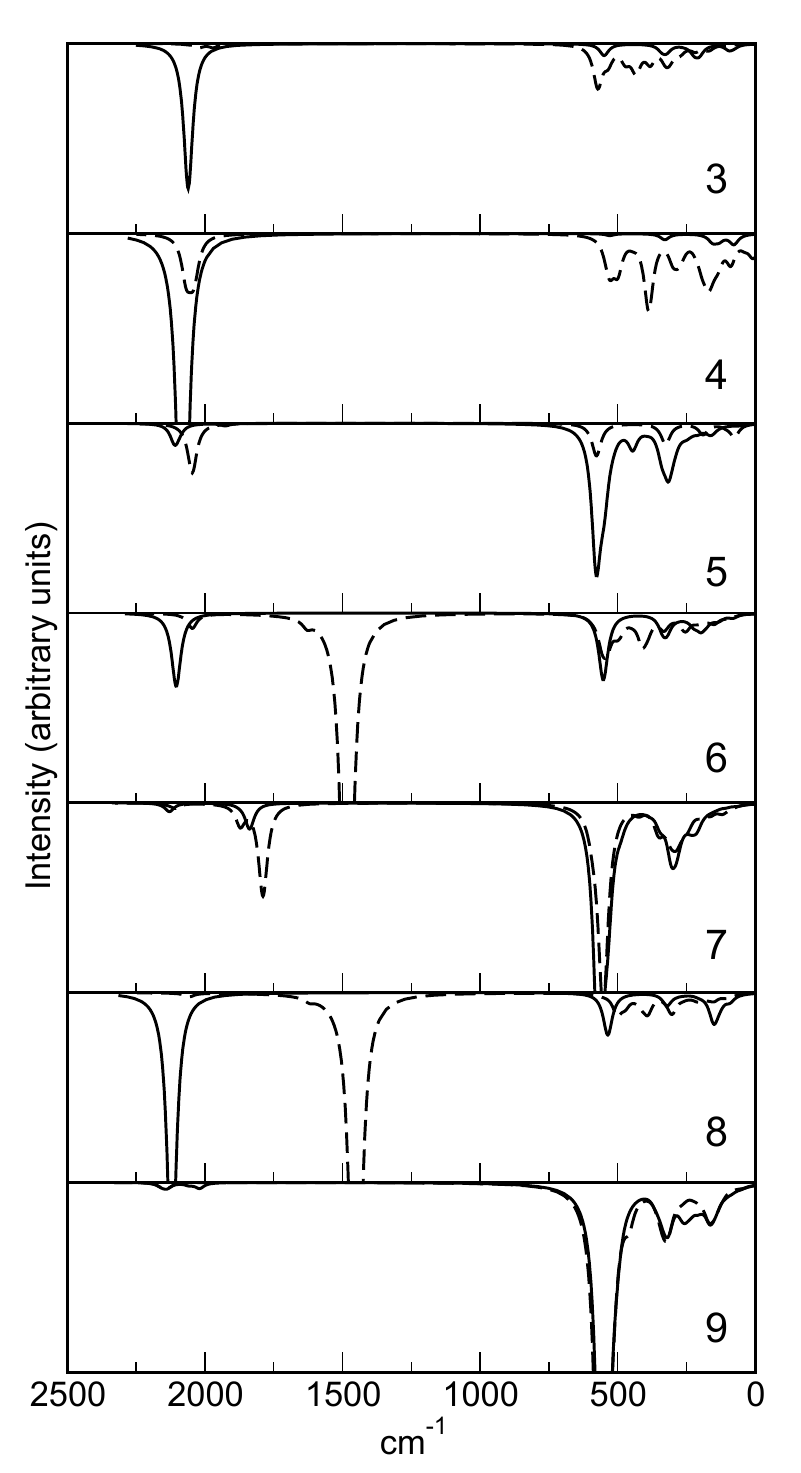}
\caption{Bismuth carbide B3LYP/Def2-TZVP IR spectra for \ce{Bi_{n}C_{2n}^+} ($3 \leq n \leq 9$).
         Solid and dashed spectra correspond to 'a' and 'b' forms respectively.}
\label{fig:IR}
\end{figure}

The calculated IR vibrational spectra for the bismuth carbides are shown in Fig \ref{fig:IR}. The chemical simplicity of these structures is evident in the small number of significant peaks in these spectra. Key vibrations observed correspond to \ce{C#C} stretches at $\sim$2,200 cm$^{-1}$ and \ce{C=C} stretches at $\sim$1,500 cm$^{-1}$ whilst more complex vibrations involving bismuth atoms and \ce{C_2} unit librations occur at much lower frequencies of around 600 cm$^{-1}$.\cite{Williams1995}

In all cases where the bonding pattern follows the simple linear scheme of Bi-C-C-Bi it was found that the \ce{C_2} units possessed a triple-bonded character. This can be clearly seen in the spectra of the simple ring-like isomers \ce{Bi_{3}C_{6}^+}(a) and \ce{Bi_{4}C_{8}^+}(a). The larger \ce{Bi_{5}C_{10}^+}(b) ring system did not follow this pattern as both the ring and cage isomers displayed triple-bond character in the \ce{C_2} vibrations. 

Compounds \ce{Bi_{6}C_{12}^+}(a) and \ce{Bi_{8}C_{16}^+}(a) also displayed \ce{C_2} unit triple-bond character which changed to reflect more double-bond character in the \ce{Bi_{6}C_{12}^+}(b) and \ce{Bi_{8}C_{16}^+}(b) structures. The similarity of the spectra for the \ce{Bi_{3}C_{6}^+}(a) / \ce{Bi_{6}C_{12}^+}(a) and \ce{Bi_{4}C_{8}^+}(a) / \ce{Bi_{8}C_{16}^+}(a) pairs is suggestive of the possibility that the heavier cage-type clusters are in fact better considered as stacked dimers of the lighter planar ring species. The \ce{C_2} unit double-bond character in \ce{Bi_{6}C_{12}^+}(b) and \ce{Bi_{4}C_{8}^+}(b) corresponds to a specific bond arrangement where a carbon atom is linked to two bismuth atoms and another carbon atom. A similar arrangement can be found for both \ce{Bi_{7}C_{14}^+} clusters, which are the only ones to display a peak at $\sim$1,800 cm$^{-1}$.

For \ce{Bi_{5}C_{10}^+}, \ce{Bi_{7}C_{14}^+} and \ce{Bi_{9}C_{18}^+} the dominant peak in the spectrum of both the a and b forms was found to be that corresponding to the combined \ce{C_2} librations and Bi-C stretches although minor contributions from the \ce{C_2} units can still be seen.

\subsection{Absorption spectra}

\begin{figure}[]
\centering
\includegraphics[width=8.5cm]{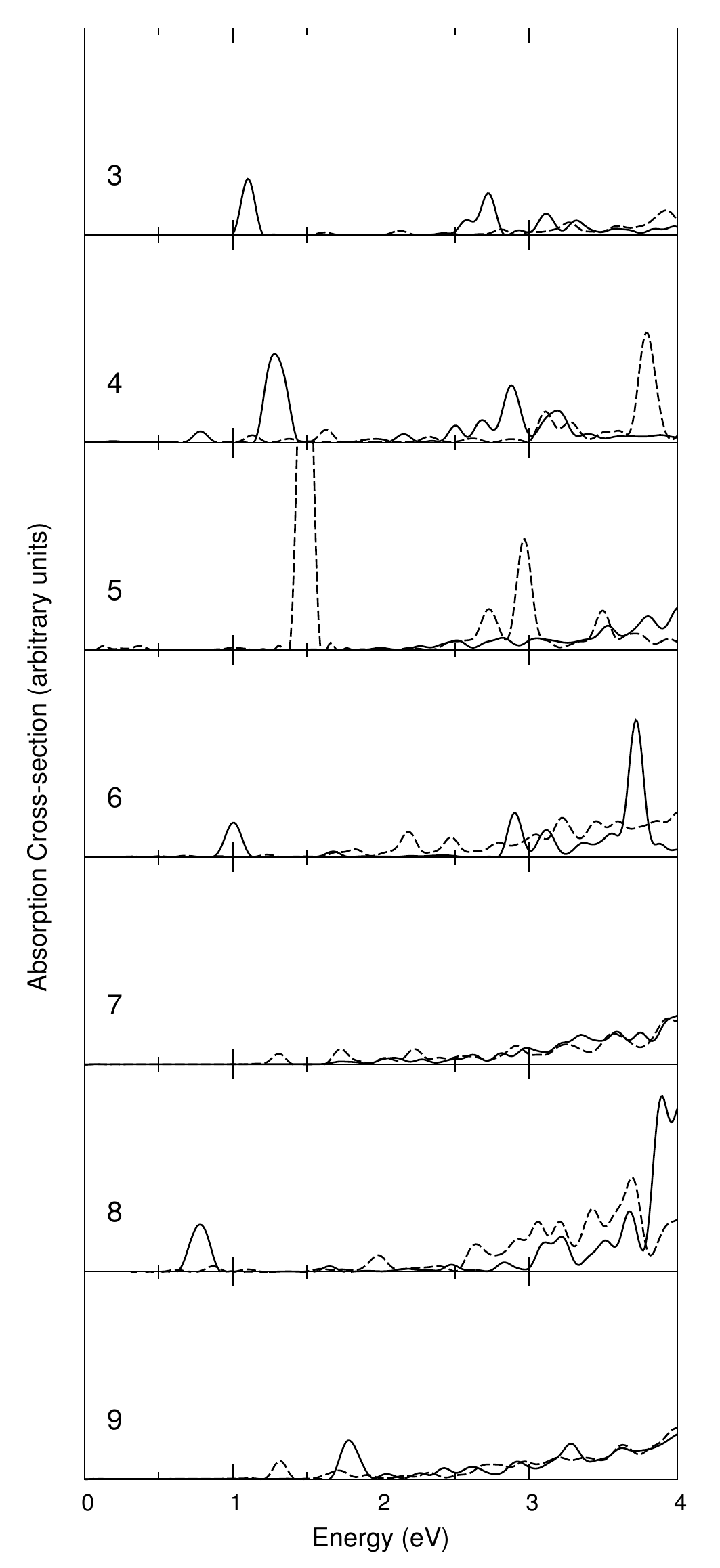}
\caption{Bismuth carbide optical absorption spectra for \ce{Bi_{n}C_{2n}^+} ($3 \leq n \leq 9$). Solid and dashed spectra correspond to 'a' and 'b' forms respectively.}
\label{fig:optics}
\end{figure}

The absorption spectra of the smaller bismuth carbide clusters were found to display features in the near infra-red (NIR) region between $\sim$0.75 and $\sim$1.75 eV that distinguished between different isomers (Fig \ref{fig:optics}). For the \ce{Bi_{n}C_{2n}^+} ($n$ = 3 -- 5) clusters this effect is very obvious with intense peaks in the NIR region being present in the ring-shaped clusters \ce{Bi_{3}C_{6}^+}(a), \ce{Bi_{4}C_{8}^+}(a), and \ce{Bi_{5}C_{10}^+}(b), but absent (or greatly diminished) in the other isomers.

To a lesser extent, a similar effect can be seen in the \ce{Bi_{n}C_{2n}^+} ($n$ = 6, 8) clusters. Despite the fact that in these cases no planar ring forms exist the NIR peaks are clearly observed for \ce{Bi_{6}C_{12}^+}(a) and \ce{Bi_{8}C_{16}^+}(a). As can be seen from the geometries in Fig \ref{fig:structures}, these clusters are effectively stacked pairs of the \ce{Bi_{3}C_{6}^+}(a) and \ce{Bi_{4}C_{8}^+}(a) rings, respectively, with Bi-Bi bonding holding the pairs of rings together. The first NIR peak energies in \ce{Bi_{3}C_{6}^+}(a) and \ce{Bi_{6}C_{12}^+}(a) are almost identical and this is also observed for the \ce{Bi_{4}C_{8}^+}(a) and \ce{Bi_{8}C_{16}^+}(a) pair. In the latter pair there is a noticeable effect where the \ce{Bi_{4}C_{8}^+}(a) peak at $\sim$1.3 eV seems to disappears. In fact, the two \ce{Bi_{4}C_{8}^+}(a) peaks below 1.5 eV correspond to the splitting of a peak by the spin-orbit interaction, which does not seem to occur in the \ce{Bi_{8}C_{16}^+}(a) cluster.

The visible region of the absorption spectra is perhaps less useful in distinguishing between the different isomers since, in most cases, there is a broad weak visible absorption increasing in the blue region. This suggests that for most of the bismuth carbide clusters some colour would be observed with the emphasis being in the red/orange range. In the cases of \ce{Bi_{3}C_{6}^+}, \ce{Bi_{4}C_{8}^+} and \ce{Bi_{5}C_{10}^+}, in addition to absorption in the blue region, the spectra display significant absorption towards the middle of the visible region for the planar ring forms but almost no absorption in this region for the other isomers. Thus, for these smaller clusters a significant difference in colour can be expected between the (a) and (b) forms.

Finally, we note that the spectra of clusters whose structures have a larger number of symmetries tend to less,better defined peaks, which is straightforward consequence of the increased number of electronic degeneracies. This effect can be clearly seen when comparing the spectra of ring and stacked ring forms with the spectra of the other isomers and is thus another feature that can help distinguishing between different clusters.

\subsection{Bonding}

In addition to the obvious structural similarities between some clusters, e.g. \ce{Bi_{3}C_{6}^+}(a) and \ce{Bi_{6}C_{12}^+}(a), the latter apparently being simply two \ce{Bi_{3}C_{6}^+}(a) rings stacked together, the calculated spectra suggested that there were strong similarities between certain clusters of different size. In an attempt to investigate the origin of this similarity a study of the bonding properties of the bismuth carbide clusters was performed. Analysis of covalent bonding as well as non-covalent interactions (NCI) was performed using approaches based on properties of the electronic change density (see Methods section above).

\begin{figure}[]
\centering
\includegraphics[scale=1.0]{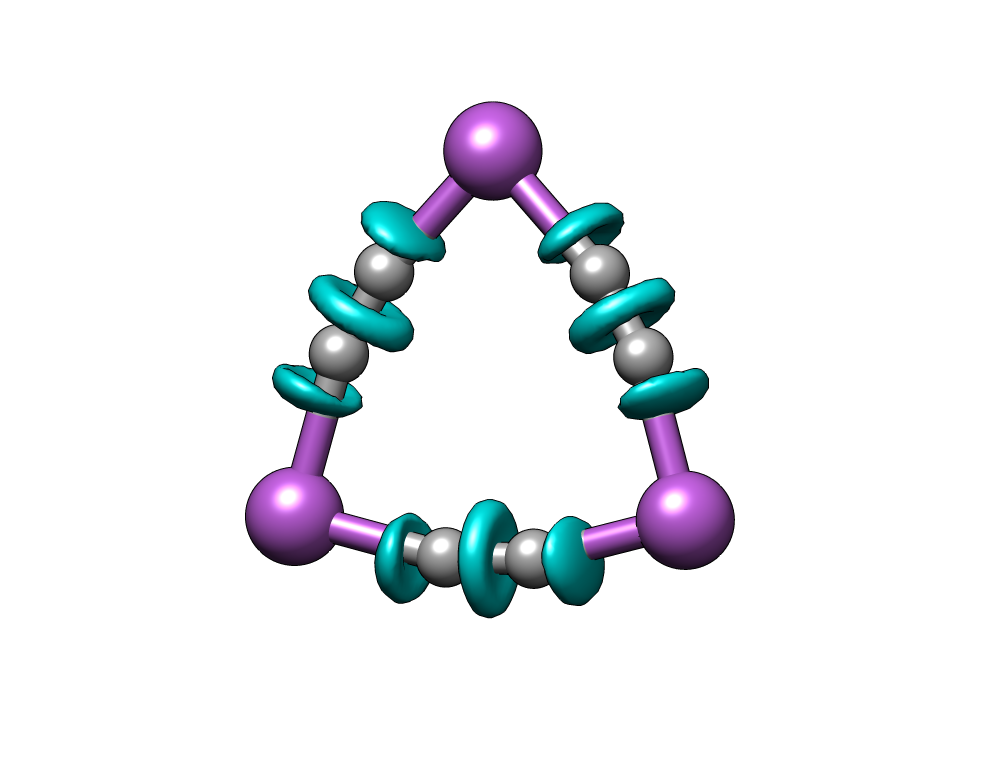}
\includegraphics[scale=1.0]{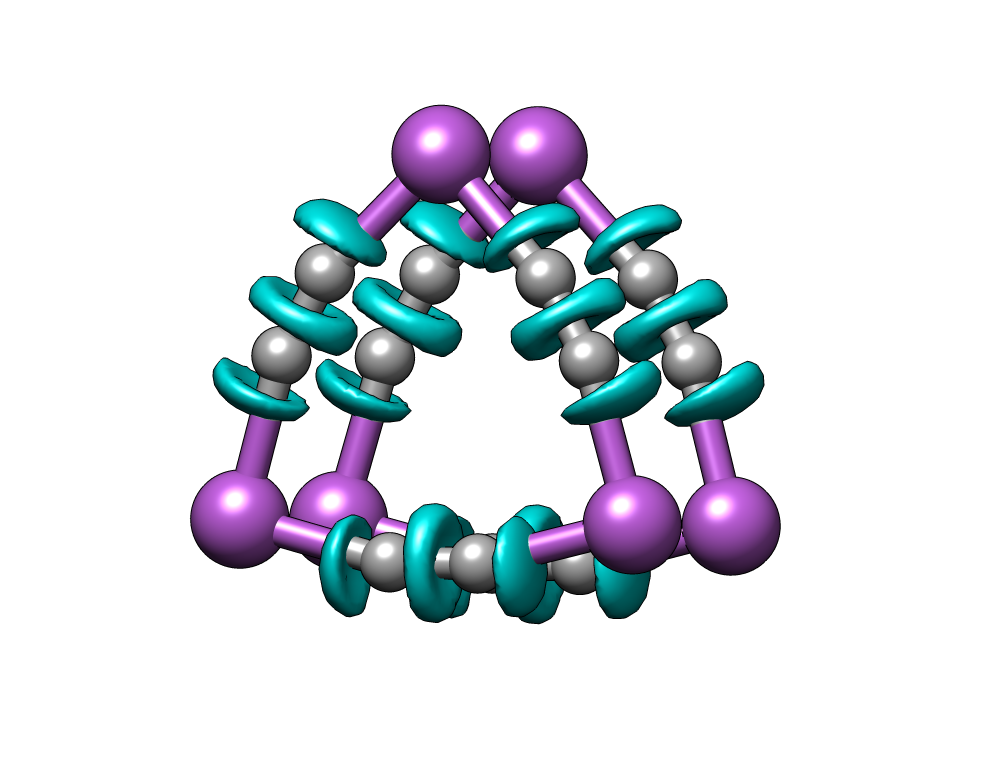}
\caption{ELF density isosurfaces (cutoff = 0.83 a.u.) for \ce{Bi_{3}C_{6}^+}(a) and \ce{Bi_{6}C_{12}^+}(a) clusters (top and bottom, respectively). Atoms: grey = C, purple = Bi.}
\label{ELF_3_6}
\end{figure}

\begin{figure}[]
\centering
\includegraphics[scale=1.0]{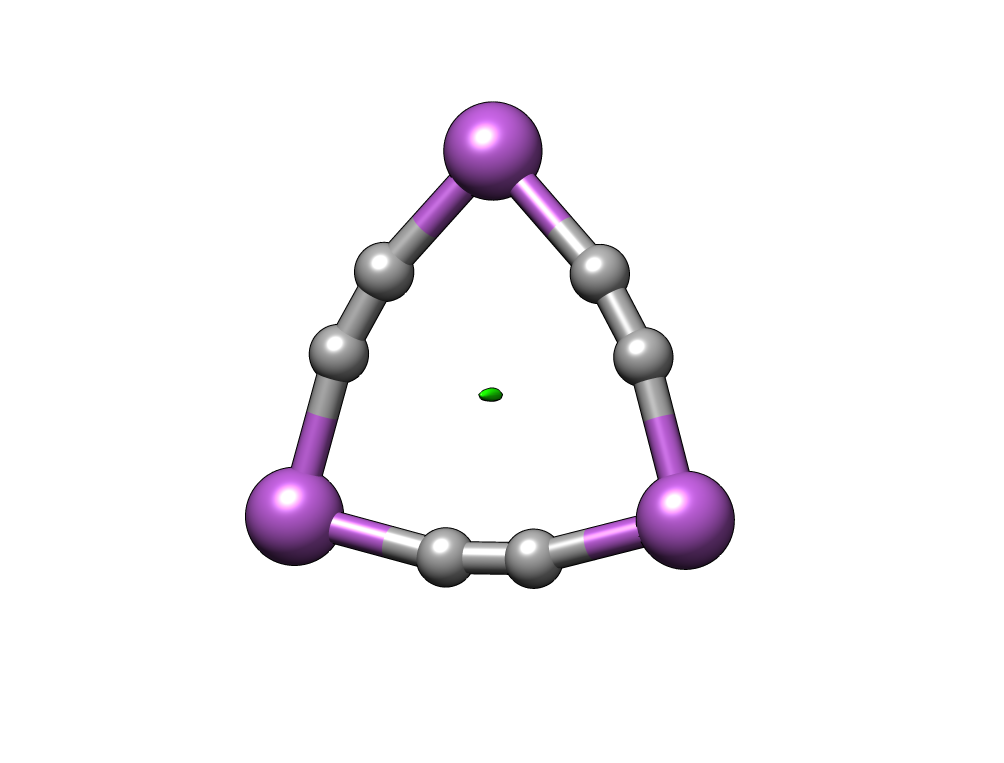}
\includegraphics[scale=1.0]{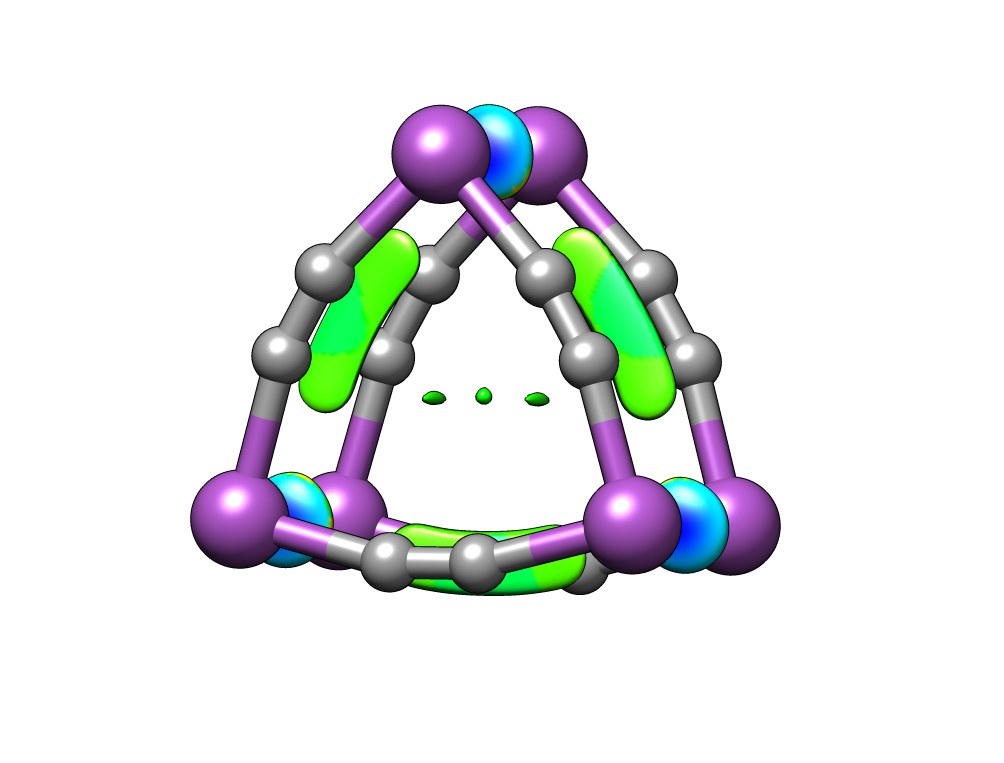}
\caption{NCI density isosurfaces (cutoff = 0.4 a.u.) for \ce{Bi_{3}C_{6}^+}(a) and \ce{Bi_{6}C_{12}^+}(a) clusters (top and bottom, respectively). Green represents moderate non-covalent interaction and blue represents strongly stabilising non-covalent interaction. Atoms: grey = C, purple = Bi.}
\label{NCI_3_6}
\end{figure}

The results of this analysis for the \ce{Bi_{3}C_{6}^+}(a) and \ce{Bi_{6}C_{12}^+}(a) clusters are shown in Figs \ref{ELF_3_6} and \ref{NCI_3_6}. The covalent bonding patterns revealed by the electron localisation function (ELF) analysis are essentially identical in the two \ce{Bi_{3}C_{6}} rings of \ce{Bi_{6}C_{12}^+}(a) to those seen in the free \ce{Bi_{3}C_{6}^+}(a) complex itself with covalent bonds within the \ce{C_2} units and between the terminal C atoms and the adjacent Bi atoms. However, the ELF analysis fails to show any concentration of electronic charge density between the stacked pairs of Bi atoms in \ce{Bi_{6}C_{12}^+}(a).

In Fig \ref{NCI_3_6} weak non-covalent interactions were observed between the stacked \ce{C_2} units (large flat green isosurfaces) suggesting that dispersion-type bonding occurs between these moieties and stabilises the complex to some extent. In contrast to the results obtained with the ELF analysis, large NCI densities were observed in the Bi internuclear regions between the stacked rings suggesting that although no covalent Bi--Bi bonding was occurring, strong non-covalent bonding between Bi atoms was responsible for stabilising this complex. In both complexes only a very small NCI volume was observed at the ring centres indicating that there is little ring strain in these systems arising from steric crowding as might be seen in smaller rings such as benzene.\cite{Contreras-Garcia2011} Similar small additional NCIs at the cage centres were observed upon formation of the \ce{Bi_{6}C_{12}^+}(a) and \ce{Bi_{8}C_{16}^+}(a) dimer systems but again do not contribute a great deal to the net stability of these systems. 

\begin{figure}[]
\centering
\includegraphics[scale=1.0]{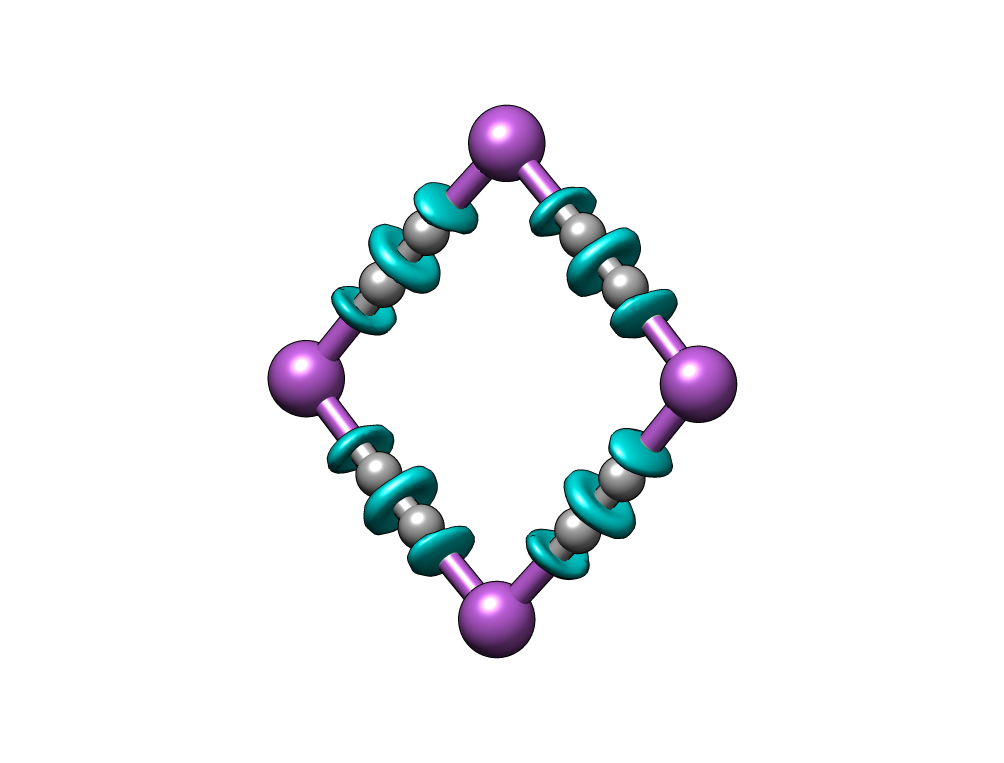}
\includegraphics[scale=1.0]{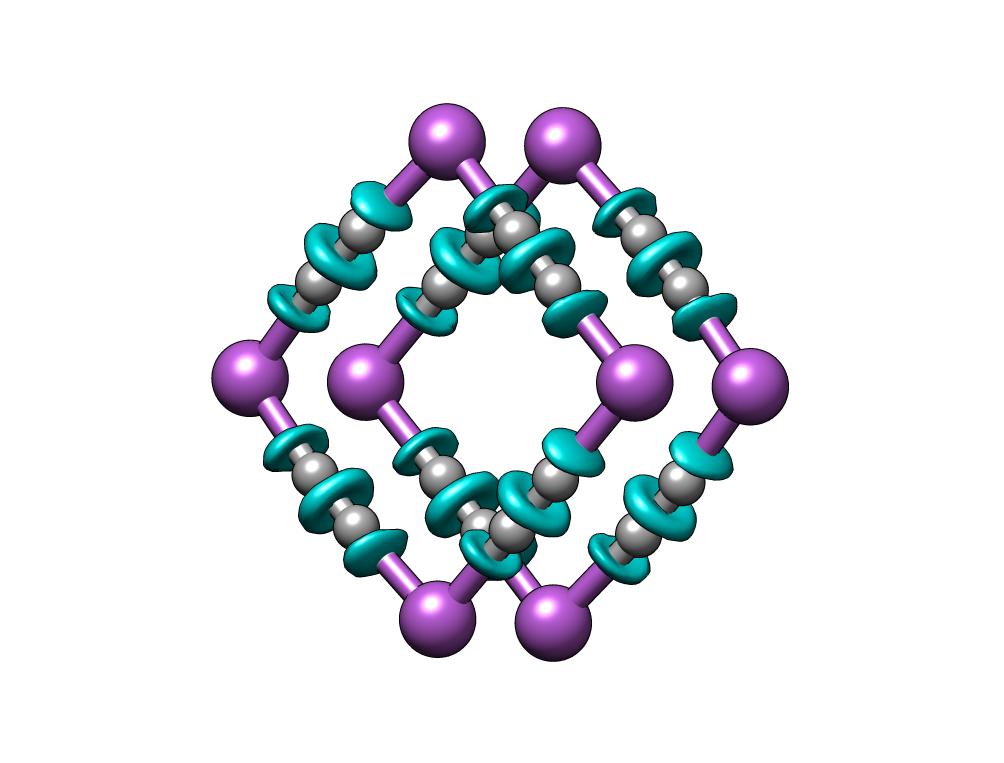}
\caption{ELF density isosurfaces (cutoff = 0.83 a.u.) for \ce{Bi_{4}C_{8}^+}(a) and \ce{Bi_{8}C_{16}^+}(a) clusters (top and bottom, respectively). Atoms: grey = C, purple = Bi.}
\label{ELF_4_8}
\end{figure}

\begin{figure}[]
\centering
\includegraphics[scale=1.0]{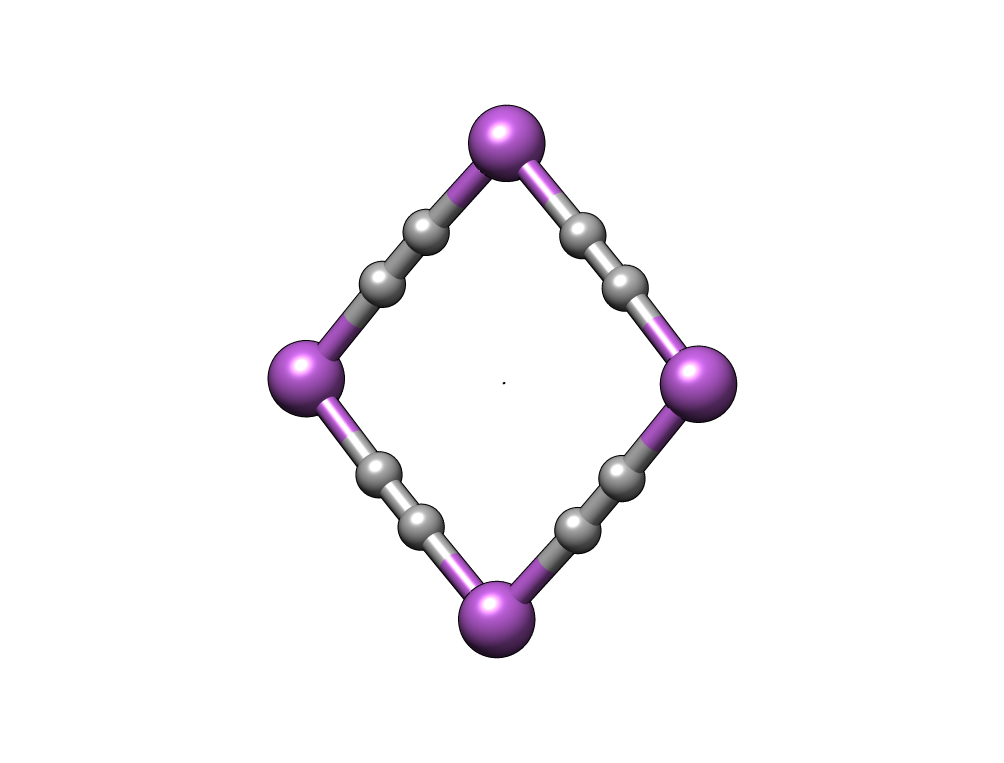}
\includegraphics[scale=1.0]{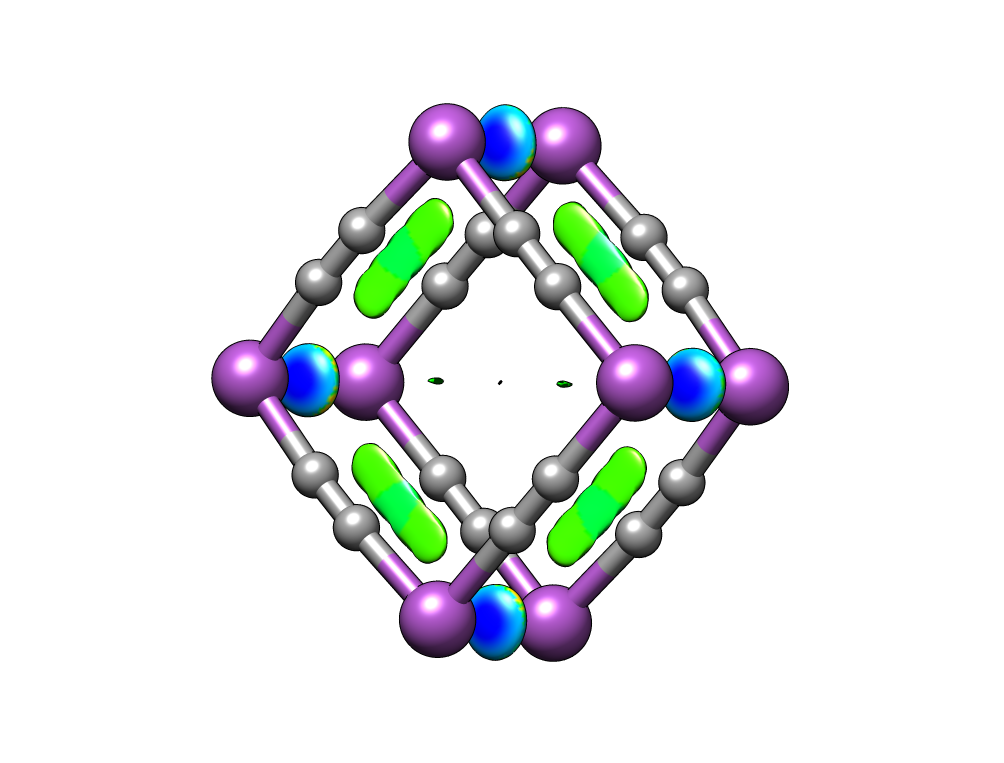}
\caption{NCI density isosurfaces (cutoff = 0.4 a.u.) for \ce{Bi_{4}C_{8}^+}(a) and \ce{Bi_{8}C_{16}^+}(a) clusters (top and bottom, respectively). Green represents moderate non-covalent interaction and blue represents strongly stabilising non-covalent interaction. Atoms: grey = C, purple = Bi.}
\label{NCI_4_8}
\end{figure}

The same bonding analyses were performed for the \ce{Bi_{4}C_{8}^+}(a) and \ce{Bi_{8}C_{16}^+}(a) complexes (Figs \ref{ELF_4_8} and \ref{NCI_4_8}). Geometrically, the larger \ce{Bi_{8}C_{16}^+}(a) complex appears to be a stacked \ce{Bi_{4}C_{8}^+}(a) dimer in the same way that \ce{Bi_{6}C_{12}^+}(a) could be formed by dimerisation of the lighter \ce{Bi_{3}C_{6}^+}(a) complex. This was borne out by the ELF and NCI results which showed, as with the previous complex pair, that covalent bonding was restricted to the ring systems and that the rings displayed weak dispersion-type interactions between the \ce{C_2} units but were dominated by stronger inter-ring non-covalent bonding between Bi atoms.

\begin{figure}[]
\centering
\includegraphics[scale=1.0]{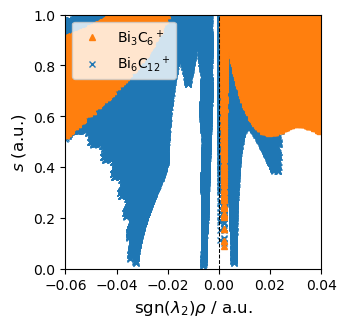}
\includegraphics[scale=1.0]{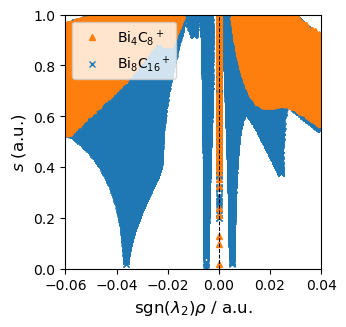}
\caption{NCI index scatter plots for the \ce{Bi_{3}C_{6}^+}(a) / \ce{Bi_{6}C_{12}^+}(a) clusters and \ce{Bi_{4}C_{8}^+}(a) / \ce{Bi_{8}C_{16}^+}(a) clusters (top and bottom, respectively).}
\label{NCI_scatter_full}
\end{figure}

\begin{figure}[]
\centering
\includegraphics[scale=1.0]{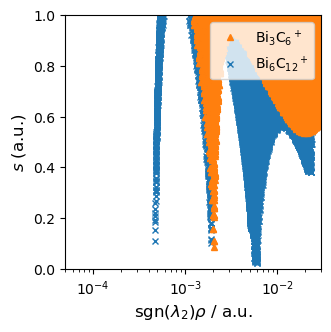}
\includegraphics[scale=1.0]{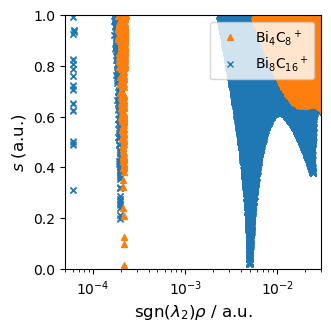}
\caption{Positive sgn($\lambda_2$)$\rho$ regions of NCI index scatter plots for the \ce{Bi_{3}C_{6}^+}(a) / \ce{Bi_{6}C_{12}^+}(a) clusters and \ce{Bi_{4}C_{8}^+}(a) / \ce{Bi_{8}C_{16}^+}(a) clusters (top and bottom, respectively). Logarithmic scale used for sgn($\lambda_2$)$\rho$ in order to highlight changes to peaks corresponding to ring strain destabilisation.}
\label{NCI_scatter_right}
\end{figure}

Scatter plots corresponding to the previously-discussed NCI isosurfaces are shown in Figure \ref{NCI_scatter_full}. The changes occurring on dimerisation are seen as an introduction of a large stabilising peak between sgn($\lambda_2$)$\rho$ = -0.03 and sgn($\lambda_2$)$\rho$ = -0.04 (corresponding to the Bi--Bi interactions) and much weaker NCIs at approximately $\pm$ 0.005 corresponding to the C--C interaction. Focussing on the right/destabilising half of the plots (Figure \ref{NCI_scatter_right}) and employing a logarithmic scale for sgn($\lambda_2$)$\rho$ it is possible to observe a slight stabilisation of the closely overlapping peaks corresponding to the ring strain in the monomer units. Similarly, the presence of a new peak closest to zero in both dimer systems is in line with the formation of small weak cage strain volumes in the isosurface plots

ELF and NCI analysis of the remainder of the complexes in Table \ref{fig:structures} can be expected to yield a similar pattern with covalent bonding in C--C and C--Bi internuclear regions but non-covalent interactions being responsible for the apparent homonuclear bonding between Bi pairs. Results for heavier and/or structurally more complicated clusters are, however, not shown here due to less easy visual interpretation of the corresponding images and isosurface plots due to increasing numbers of NCIs and lowering of the symmetry of these systems. 

The data obtained in the current work help to rationalise the similarities in the calculated spectra of certain pairs of complexes such as \ce{Bi_{3}C_{6}^+}(a) and \ce{Bi_{6}C_{12}^+}(a) or \ce{Bi_{4}C_{8}^+}(a) and \ce{Bi_{8}C_{16}^+}(a) by showing that in fact the heavier complexes are essentially non-covalent dimers of their lighter counterparts and might therefore be expected to behave very similarly from the point of view of spectroscopic analysis. This might also go some way to explain the lack of these heavier complexes in the product mixture obtained from the gas-aggregation experiments; if these are indeed dimers bound by non-covalent interactions then it is not unreasonable to expect them to be less abundant than the corresponding monomer units. 

\section{Conclusion}

First-principles calculations at the (time-dependent) density functional theory ((TD)DFT) level have been performed in order to investigate the spectroscopic characteristics of the cationic bismuth carbide clusters \ce{Bi_{n}C_{2n}^+} ($3 \leq n \leq 9$). The results of these calculations show that using a combination of UV-Vis absorption spectroscopy and IR vibrational spectroscopy it should be easy to distinguish between cluster isomers of identical mass.

The presence of an intense peak in the vibrational spectra at $\sim$2000 cm$^{-1}$ is indicative of the presence of the \ce{Bi_{n}C_{2n}^+}(a)(n = 3, 4, 6, or 8), while an intense peak at $\sim$1500 cm$^{-1}$ is indicative of the presence of either the \ce{Bi_{6}C_{12}^+}(b) or \ce{Bi_{8}C_{16}^+}(b) clusters. Absorption spectroscopy highlights some important differences in the optical response of the lighter clusters that were seen to be most abundant in the gas-aggregation source experiments. In particular, \ce{Bi_{3}C_{6}^+}(a) and \ce{Bi_{4}C_{8}^+}(a) display significant peaks between 1 and 1.5 eV and between 2.5 and 3 eV.

The charge density based bonding analysis performed in the present work suggests that many of the larger clusters may in fact owe a lot of their calculated structural stability to non-covalent interactions as opposed to covalent bonds. A similar observation has been made in layered crystals of Bi/Se alloys where strong non-covalent Bi--Bi interactions are seen between layers in the crystal structure.\cite{Christian2015} This is exemplified in the current work by the \ce{Bi_{6}C_{12}^+}(a) and \ce{Bi_{8}C_{16}^+}(a) clusters which the present analyses suggest are in fact best described as non-covalently bound dimers of the corresponding \ce{Bi_{3}C_{6}^+}(a) and \ce{Bi_{4}C_{8}^+}(a) planar ring species, albeit with an extra electron in the dimeric case in order to maintain the unit positive charge. The open-shell nature of these clusters complicates the bonding picture but it is likely that a mechanism similar to the $n \rightarrow \pi^{\ast}$ non-covalent bonding observed in X--arene systems (X = Sb, Bi).\cite{Caracelli2013,Sing2015} For this reason it is expected that although they may be expected to be thermodynamically stable based on the results of these and previous calculations, at finite temperature kinetic instability may lead to their decomposition. This is consistent with the mass distribution for these clusters observed experimentally.\cite{Yamada2009} 

Studies of the temperature-dependent dynamic behaviour of the clusters are planned in order to investigate further the extent to which the non-covalent bonding observed here may contribute to the stability (or otherwise) of these systems. In addition, ongoing developments of the PyNCI software to allow better isolation of individual regions of interaction may shed further light on the more complicated systems and their behaviour.

\acknowledgement

This work was supported by national funds from FCT – Funda\c{c}\~{a}o para a Ci\^{e}ncia e a Tecnologia, I.P., within the projects UIDB/04564/2020, UIDP/04564/2020, PTDC/FIS/103587/2008, SFRH/BPD/44608/2008, CONT{\_}DOUT/11/UC/405/10150/18/2008, POCI-01-0145-FEDER-032229 and CENTRO-01-0145-FEDER-000014, 2017-2020. The authors thank the Laboratory for Advanced Computation of the University of Coimbra for the provision of computer resources, technical support and assistance. BFM acknowledges support from the Donostia International Physics Centre and the Centro de F\'{i}sica de Materiales (UPV/EHU), San Sebasti\'{a}n, Spain.


\bibliography{bismuth_carbide}

\end{document}